**Nanoscale detection of metastable states in porous and granular media**

Eduard Ilin[1,2], Yaofa Li[3], Eugene V. Colla[1], Kenneth T. Christensen[3], Muhammad Sahimi[5], Maxim Marchevsky[6], Scott M. Frailey[4], and Alexey Bezryadin[1]

[1] Department of Physics, University of Illinois at Urbana-Champaign, Urbana, IL 61801, USA
[2] Far-Eastern Federal University, Vladivostok, 690090, Russia
[3] University of Notre Dame, Notre Dame, IN 46556, USA
[4] Illinois State Geological Survey, Prairie Research Institute, University of Illinois at Urbana-Champaign, Champaign, IL 61820, USA
[5] Mork Family Department of Chemical Engineering and Materials Science, University of Southern California, Los Angeles, CA 90089-1211, USA
[6] Lawrence Berkeley National Laboratory, Berkley, CA 94720, USA

*Microseismicity in subsurface geologic environments, such as sandstone gas reservoirs, is expected in the presence of liquid or gas injection. Although difficult to predict, the potential for microseismic events is important to field-scale projects, such as geologic storage of $CO_2$ whereby the gas is injected into natural sandstone formations. We conjecture that a primary factor causing microseismicity is the existence of metastable states in granular porous medium and provide experimental evidence for its validity. External perturbation trigger abrupt relaxation events, which, with a certain probability, can grow into macroscopic microseismic events. Here the triggering perturbation is produced by cooling to a cryogenic temperature. As the "sensor" for the abrupt relaxation events we use thin Al films deposited on the sandstone surface. We show that as the temperature is varied, the films' resistance exhibits sharp jumps, which we attribute to mechanical restructuring or microfractures in the fabric of the sandstone. We checked the superconducting characteristics of the Al thin films on the sandstone and found microwave-induced Shapiro steps on the voltage-current diagrams. Such quantized steps provide indicates that the film is made of a network of nanobridges, which makes it ever more sensitive to abrupt relaxation events occurring in the substrate, i.e., in the underlying sandstone.*

## I. INTRODUCTION

Extraction of natural resources from the subsurface and industrial processes that requires injection of fluids into geologic formations change the pore pressure and stress field near the injection well. In the presence of certain geologic features, naturally occurring stored energy can be released in the form of seismicity, i.e. earthquakes [1,2]. Over the past 10 years, earthquakes have been uniquely correlated with large-volume water injection in several oil- and gas-producing states (e.g., in Oklahoma, Ohio and Texas) [3,4]. To ensure safe usage of such natural resources, the industry, the government and the public have demonstrated a keen interest in advancing the ability to control seismicity. Yet, understanding of microseismicity that results from fluid injection remains incomplete, with microseismicity defined as energy release of an event of magnitude (M) of −2 to 0, which are generally low energy that are not felt by humans on the Earth's surface, whereas the nanoseismicity range is M −4 and −2 [5].

Another use of subsurface pore space that may cause injection-induced seismicity is related to humans' attempt to mitigate the Earth's increasing atmospheric concentration of $CO_2$. To slow global climate change, many countries have been considering the geologic storage of $CO_2$ to reduce the emission of greenhouse gases into the atmosphere. Various studies have been completed to determine the $CO_2$ storage capacity available in geologic formations [6]. Most of such studies have concluded that adequate storage space will be available for decades, while the world transitions to lower emission technologies. However, to have a meaningful effect on the Earth's temperature, an unprecedented volume of fluid needs to be injected in the subsurface, and many researchers and policy makers expect the seismicity associated with $CO_2$ injection to be a major challenge to commercial-scale storage. Laboratory experiments showed that if a fluid is released from pores of geological materials then layers of the material begin to slide with respect to each other and low frequency oscillations can be detected, similar to those detected in long term volcanic activity [7]. The reverse process, namely the process of fluid injection, can also generate seismic oscillations, as was shown by Turkaya et al. [8]. At present, however, microseismicity has only been recorded in pilot-scale and demonstration of $CO_2$ storage sites.

The key to controlling injection-induced seismicity is understanding the physical mechanisms by which it occurs. Although much of the literature has focused on large-scale features, such as geologic faults that are a result of tectonic activity within the Earth's crust [9], all the energy release must begin at the scale of micro- and nano-size pores of rock. The displacement scales of micro- and nanoseismicity are, respectively, 0.4 – 4 mm and 40 – 400 µm. Note that, as defined, the displacement scale of nanoseismic events is comparable to the size of one sand particle. On the other hand, low-energy releases, which propagate on a scale of thousands of sand particles, may be precursors to the events felt on the Earth's surface. But in the samples used, where the sample size is decimetric, the slip can be of the order of 40 μm, with a magnitude of around −4 or less. Note that there is always a much higher rate of small displacement events in which only one or a few sand grains are involved, yet only a small fraction of them can grow into a large-scale seismic event, such as an earthquake. In addition, the bond between sand particles is usually much weaker than the connection between the molecules within the particles. Thus, any fracture should develop in the space between the sand particles, i.e., in the cementation material linking the grains together. Identifying and measuring local fractures at the scale of individual sand particles may lead to a transformative understanding of the causes of seismicity in the subsurface, which will lead to safe usage of its resources. Our guiding principle here is that each earthquake begins with a displacement of a single grain. Thus, understanding the conditions at which single grains rearrange their mutual positions is important.

Typically, in laboratory experiments, geological samples are usually studied by generating internal stress, associated with external mechanical perturbations. This causes micro-shifts and micro-sliding events, which are detected by sensitive acoustical techniques [10,11]. Our approach is to use strong temperature variations instead of mechanical deformations. Temperatures variations can also induce internal stress as well as shear strains, especially if the temperature is changing nonuniformly, which is the case here.

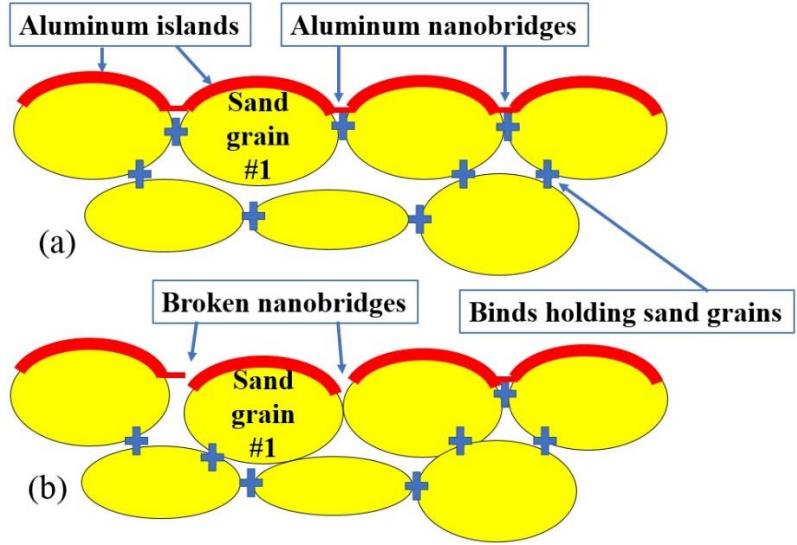

*Figure 1.* *(a) Schematic diagram of the metal-coated sandstone sample. The sand particles, shown as yellow oval shapes, are held or bonded together by cementation material ("bonds"), shown schematically by blue crosses. The Al film deposited on the surface of the sample by thermal evaporation is shown by thick red lines. The film covers the surface and is used for making electrical measurements. The metallic film is made of Al islands, which are electrically connected to one another by nanobridges (short horizontal red lines) at the points where sand particles touch, i.e., the nanobridges (thin red lines) are located above the bonds. (b) If a bond between two sand particles breaks, then the corresponding Al nanobridge is also interrupted. In this illustration, sand particle #1 is shifted downward by some external perturbation, such as reducing the temperature of the sample, which causes thermal contraction. The shift breaks the nanobridges linking the Al island on top of the particle to its neighbors, thus increasing the sample's electrical resistance.*

To observe such local fractures involving only a single or a few sand grains, we have developed a sensitive technique: The sandstone samples were coated with a nanometer-scale thin metallic Al film, as is illustrated in Figure 1(a). The resulting thin Al film is composed of "islands" that form on the surface of individual sand grains. The Al islands so obtained are connected through Al nanoscale constrictions or "nanobridges" that are naturally located between the sand particles, namely, over the bond regions between the particles. Thus, if a microscopic fault appears locally, i.e., if a bond between two sand particles breaks, it has a tendency to rupture the corresponding Al bridge, as shown schematically in Figure 1(b). Local bonds can break because the positions of some local configurations of particles may not be in equilibrium, or, in other words, may be metastable [12,13,14].

In this paper, we describe measurements of electrical resistance on a sample of centimeter-scale natural rock, the Berea sandstone, and the changes in the deposited film resistance induced by temperature variations. We observed sharp jumps in the resistance, which were due to breaking of the nanobridges present in the Al film. Control experiments on rough Si substrates with surface pores, which did not have a granular texture, did not exhibit such resistance jumps. Therefore, we conjecture that the observed strong upward resistance jumps are related to breaking of the local bonds between sand particles, which causes some nanobridges to break. Multiple repetitions of the cooling–warming cycle, i.e., the thermocycling, greatly reduced the number of resistance jumps, presumably because the thermocycling helps the existing metastable nonequilibrium states to relax. This is similar to the cases where a cycling mechanical stress was applied, which caused a reduction of the number of internal metastable increased-stress states. Thus, mechanical cycling led to a decrease of microseismic activity [15]. We observe an analogous effect with thermal cycling.

The rest of this paper is organized as follows. In the next section we describe fabrication of the samples used in the experiments. Section III describes the measurements. The main results are presented in Sec. IV. To provide further evidence for the validity of our working hypothesis, we carried out acoustic emission measurements, whose results are presented in Sec. V. The paper is summarized in Sec. VI.

## II. SAMPLE FABRICATION

The Al evaporation method has previously been used to manufacture such products as Al mirrors [16]. Aluminum is a convenient material for making thin films because it does not change even when exposed to air. Thin films of Al also become superconducting at low temperatures [17,18]. Aluminum has been used for fabricating nanowires and nanobridges with diameters of only a few nanometers that are superconducting at low temperatures [19,20].

Our samples were prepared by thermal evaporation of Al film (99.99% pure) on variable substrates under vacuum ($10^{-5}$ bar). The Al films have been deposited using a thermal evaporation commercial equipment, Denton Vacuum DV-502A. The machine is equipped with a film thickness monitor Model TM-350, MAXTEK, INC. It operates by measuring the resonance frequency of a quartz resonator and calculating the amount of materials deposited based on the frequency shift.

Aluminum ingots were placed on a tungsten crucible, which was heated above the boiling point of Al. Thus, Al began to evaporate and condense on the sandstone sample, covering it with a thin film of aluminum. The rate of deposition was 3.0 - 4.0 nm/s. We used three types of substrates: (1) sandstone plates, which were approximately 1.5 - 2.0 mm thick; (2) surface-rough, etched single-crystal Si substrates having a simulated porosity of $P = 0.46$, and (3) nonporous polished Si wafer chips covered with silicon oxide. For the two-dimensional (2D) porous media used herein, the porosity is defined as the ratio of the etched area to the total area of the sample following previous studies [21,22], and its actual value was measured using a 3D confocal laser microscope Olympus OLS 4000.

The sandstone substrates were cored from a block of Berea sandstone. This rock, quarried from Berea, Ohio, is a fine-grained sandstone composed mainly (~ 95% of solid volume) of sub-rounded to rounded quartz grains. Other constituent minerals include kaolinite (2%), microcline (1.5%), and muscovite (1.2%). Both the surface-etched Si and the flat, polished Si chips were fabricated from a 0.5 mm thick single-side-polished silicon wafer. To prepare the surface-etched Si substrate, we first generated a 2D pattern, which was inspired by a previous study [23] and represents a reprint of the pore structure of real sandstone. The pattern was transferred to a bare silicon chip by using standard photolithography. The chip was then etched by using inductively coupled plasma-deep reactive ion etching to a depth of approximately 30 µm [24,25]. To etch the Si, a mixture of two gases, $SF_6$ and $C_4F_8$, was used in 5s - 2 s cycles. The flow rates for the $SF_6$ and $C_4F_8$ were 300 and 130 sccm (standard cubic centimeters per minute), respectively. The radio frequency source power was 80 W. Finally, the chip was diced to the proper size and cleaned sequentially by using acetone, methanol, water and $O_2$ plasma to remove the residual photoresist. To create a uniform silicon oxide layer on substrate (3), the silicon chip was oxidized at 1100°C for 60 min in a dry oxygen furnace, so that approximately 100 nm thick $SiO_2$ was grown on the surface [25]. The metallic Al film was 530 nm thick for all three types of substrates. Typically, the lateral dimensions of the metallic film were $2 \times 6$ mm (which are slightly smaller than the substrate size). The electrical contacts were gold wires connected to the Al film by indium spheres ("In dots"), shown in Figure 2. Indium is a soft metal that adheres to the substrates covered by Al films and makes good electrical connections. The distance between the closest pairs of In dots was approximately 0.5 – 1 mm, whereas the distance between the farthest dots was approximately 5 mm. Figure 2 shows photographs of a part of each sample.

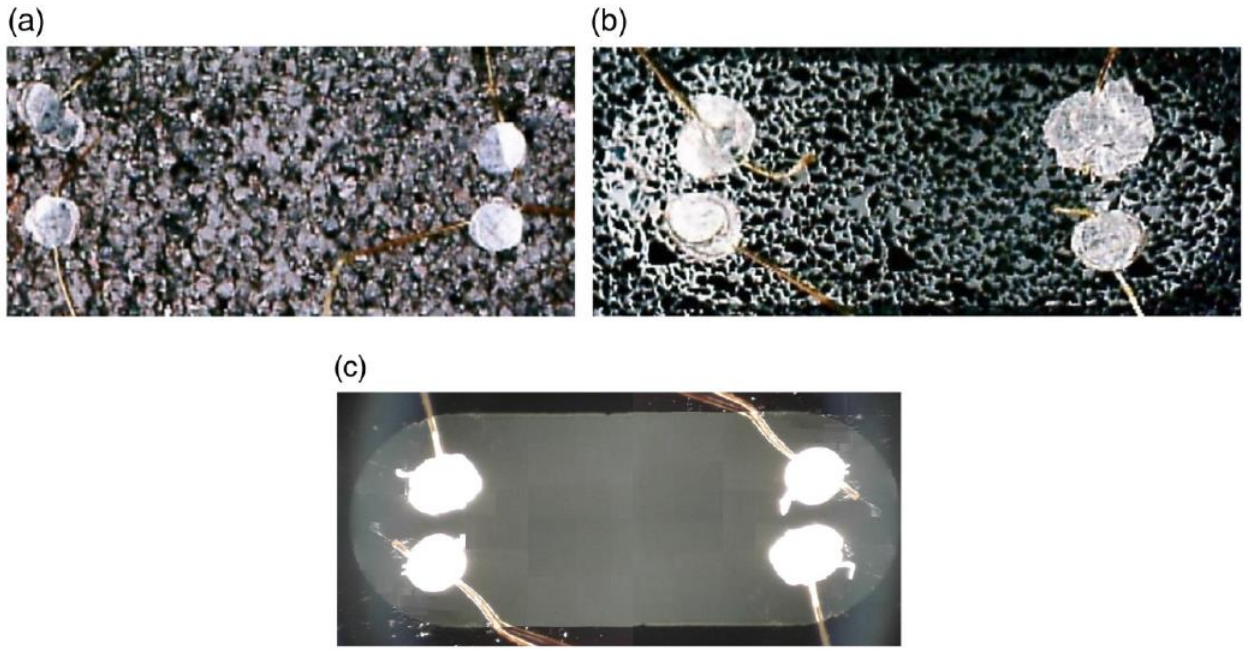

*Figure 2. Photographs of the three types of samples. (a) Granular Berea sandstone coated with a 530 nm thick Al film. Four In dots were placed on the sample to achieve electrical contacts, with each dot being a sphere of metallic In that can be squeezed onto the surface of the sample to establish electrical contact. The diameter of the spheres is approximately 250 µm. The gold wires connected to the In dots are also visible. (b) Non-granular surface-roughened Si substrate with pores etched into a single-crystal Si wafer, also coated with 530 nm thick Al film. (c) A flat, non-granular single-crystal Si wafer (oxidized on the surface), coated with 530 nm thick Al film.*

## III. MEASUREMENTS

The resistance of the Al films was measured by the two-probe methods - see Figure 3(a) - as well as by the four-probe method shown in Figure 3(b), using electric circuits of Fig.3. The four-probe measurements were done using a Keithley 2000 instrument, designed for this sort of electrical measurements. We measured the resistance of the Al films as a function of time, temperature, or both, as the sample was cooled down or warmed up. The sample was mounted on a holder and then placed into a dipstick, which was a hermetically sealed stainless-steel pipe. Before the cool-down, air was pumped from the dipstick to $10^{-3}$ bar and a small amount of He gas was injected. The role of the He gas was to ensure a good thermal exchange between the walls of the dipstick, which were cold because they made direct contact with the liquid nitrogen, and the sample, hence serving as a thermal exchange gas. To begin our standard cool-down process, the dipstick was placed into a Dewar with a cryogenic fluid, namely liquid nitrogen. Measurements were then begun immediately. The temperature gradually dropped, over approximately 10 min, to the base temperature, which was 77 K in this case. To begin the warm-up process, the dipstick was removed from the liquid nitrogen bath and gradually warmed to room temperature. Thus, the dipstick (and not the sample) was exposed to the moisture of ambient air during the warming up stage, so the sample always remained dry. In the warming-up experiments the electrical resistance measurement begins immediately after the dipstick is removed from the liquid nitrogen. The warm-up process continued for approximately 10 min.

### A. Type I measurements

In Type I measurements, we used the two-probe method shown in Figure 3(a). In this case, the ohmmeter (Keithley 2000) was attached to two contacts (i.e. two In dots) on the sample. The device was initiated in the resistance measurement mode "Ω2", which symbolizes the two-probe measurement. The dipstick with the sample was placed into the dewar with liquid nitrogen, and then the LabView computer program controlling the measurement was launched. The resistance was measured between pairs of contacts, either *a-c* or *b-d,* as labelled in figure 3(a). During the measurement process, switch 3 was periodically switched between the pairs *a-c* and *b-d* at intervals of roughly 3 – 5 s. Thus, we effectively obtained two sets of data for the time-dependence of the sample resistance, one for each pair of contacts (*a-c* and *b-d*), as illustrated in Figure 4(a). We will discuss this in detail later in the Results section.

The main purpose of this measurement was to reveal changes, especially abrupt jumps, in the film's resistance. The rock itself was insulating. The Al film adheres well to the rock surface. Any mechanical changes in the rock must cause some corresponding changes in the geometry and, therefore, in the electrical resistance of the Al film deposited on the rock. When the two probe-measurements were performed, the contact resistance might also have had an impact, and by alternating the measurement between pairs *a-c* and *b-d* we were able to separate the causes of the change in the resistance. Specifically, if one of the contacts (In dots) partially detached from the film, then an abrupt change would be seen in only one pair of contacts, either *a-c* or *b-d*. If the resistance jump was associated with some process occurring within the Al film, however, then a resistance jump would occur in both sets of data (*a-c* and *b-d*). Such results are discussed below in the Results section. For example, Figure 4 represents the results for the resistance measurements for three qualitatively different samples with different substrates measured during our standard cool-down process. Figure 5 shows the time-dependence of the resistance for the same samples in our warm-up process.

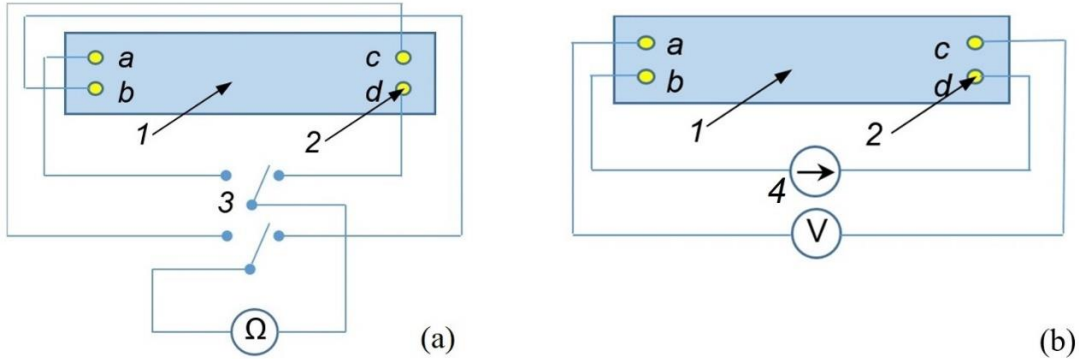

*Figure 3.* Measurement schemes: (a) Two-probe measurement. 1: Al film; 2: In dot; 3: switch; Ω: ohmmeter (Keithley 2000). (b) Four-probe measurement. 1: Al film; 2: In dot; 4: source of current; V: voltmeter (Keithley 2000).

### B. Type II measurements

In Type II experiments, we used the four-probe resistance measurement method, as shown in Figure 3(b) [26]. The four-probe technique allowed us to eliminate the contribution of contact resistance from the overall measured resistance. In this approach a current was applied through one pair of contacts, i.g., the *b-d* pair, and the voltage was measured on the other pair, i.g., the *a-c* pair. The sample's resistance was then obtained by dividing the measured voltage by the value of the applied current. Because there is no current through an ideal voltmeter, we observed no additional voltage drop on the voltage contacts (*a* and *c*). Therefore, the voltage measured was the same as the voltage drop in the film, excluding the contact. Thus, the resistance of the In contacts was not included in the measured resistance. The four-probe measurement was done using a Keithley 2000 instrument, which was initiated in the "Ω4" mode. In some cases, the four-probe measurement was done by sweeping the current and measuring the voltage at a high rate, with many points per cycle. The resistance was then obtained by measuring the slope of the resulting voltage–current curve [27]. The results did not show any dependence on the method employed.

### IV. RESULTS

Figure 4 shows how the resistance of the samples evolved during a typical cool-down. The resistance measured represents that of the thin Al films deposited on different substrates, because the substrates were insulating in all the experiments. We plot the resistance versus time because the temperature decreased with time. Immediately before the measurements began, the dipstick containing the sample was immersed in liquid nitrogen, so that the temperature decreased monotonically with time, from room temperature to that of liquid nitrogen, 77 K. The graph shown in Figure 4(a) corresponds to the sandstone sample. In this case, as the temperature decreased, the resistance of the Al film decreased slightly at the beginning of the process, but then increased sharply at times after $t \sim 100$ s. We characterize the changes as smooth increases as well as jumps, marked by the arrows. The jumps are defined as instantaneous changes of the resistance of an amplitude three times larger than the average noise amplitude of the setup. Overall though, the resistance increased by up to 30%. When the temperature approached that of liquid nitrogen at 77 K, the jumps disappeared because the temperature changed very slowly as the system approached its new equilibrium state. Note that Figure 4(a) has two graphs. As explained in the Measurements section, the two lines correspond to the two pairs of contacts, namely *a-c* [red curve (*a*)] and *b-d* [blue curve (*b*)]. This is a Type I measurement, as defined above. The jumps shown by the arrows coincide on both measurements [red curve (*a*) and blue curve (*b*)], which provides strong evidence that they originated from the Al film and not from the In contacts.

The results in Figures 4(b) and (c) represent the two controlled samples, namely, the Si substrate with simulated porosity and roughness, shown in Figure 4(b), and a flat, polished Si substrate depicted in Figure 4(c). The thickness of the Al film on the surface was the same, 530 nm. Both control samples exhibited a decrease in their resistance when the temperature decreased, which provides strong evidence that breaks or cracks did not occur in the control samples during the cool-down, unlike in the Al film deposited on sandstone. Such a decrease in the resistance with cooling (Fig.4b,c) is typical for ordinary metals [28] and can be explained by the fact that the density of phonons decreases as the temperature drops. Phonons can scatter electrons, thus increasing the resistance at higher temperatures, at which point they have a greater density. In the control samples, we did not see jumps in the resistance curves shown in Figures 4(b) and (c), which indicates that the Al film did not break with cooling by itself, even in the sample fabricated on the Si substrate with induced artificial roughness. Moreover, the resistance of the sample with etched Si substrate was very similar to that of the sample formed on the perfectly flat Si substrate. Only very small jumps were observed on the etched Si sample shown in Figure 4(b), but they did not occur on both pairs of contacts at the same time. Thus, they are probably related to the changes in the contact resistance and not in the tested Al film. It should be stressed that there is a large difference between the measurements with the control samples and the sandstone samples: the resistance of the film on the sandstone substrates increased with cooling, whereas the resistance of the films on the monocrystalline Si substrate decreased, even when the simulated surface pores and surface roughness were present. This fact can be understood by making a natural assumption that sand grains shift with cooling, thus causing a large number of very small breaks or cracks to develop in the Al film placed on the sandstone. Larger jumps of the resistance indicate strong shifts of the sand grains from their original positions.

We, therefore, suggest that the sandstone is inherently unstable because of its granularity. It is true that compression and shear mechanisms are dominant when microseismicity is concerned. In our case the cooling process causes a certain compression also. In addition, different areas might cool with different rates that cause inhomogeneous compression and, correspondingly, shear deformations. The film on the sandstone contains many nanoscale bridges or weak links (Fig.1), which limit the conductivity and can break easily if the underlying sand grains shift from their initial position because of relaxation of metastable regions. Because of such weak links, the resistance of the sandstone-supported film shown in Figure 4(a) is roughly 10 times larger than that of the Al film on the control samples in Figures 4(b) and (c). The existence of a small number of nanobridges was confirmed by the superconductivity measurements and the observation of quantum steps induced by microwave radiation, as discussed below. The fact that the resistance of the film goes up with cooling probably indicates that there is a large number of small shifts of the surface grains which produce micro-cracks in the top Al film.

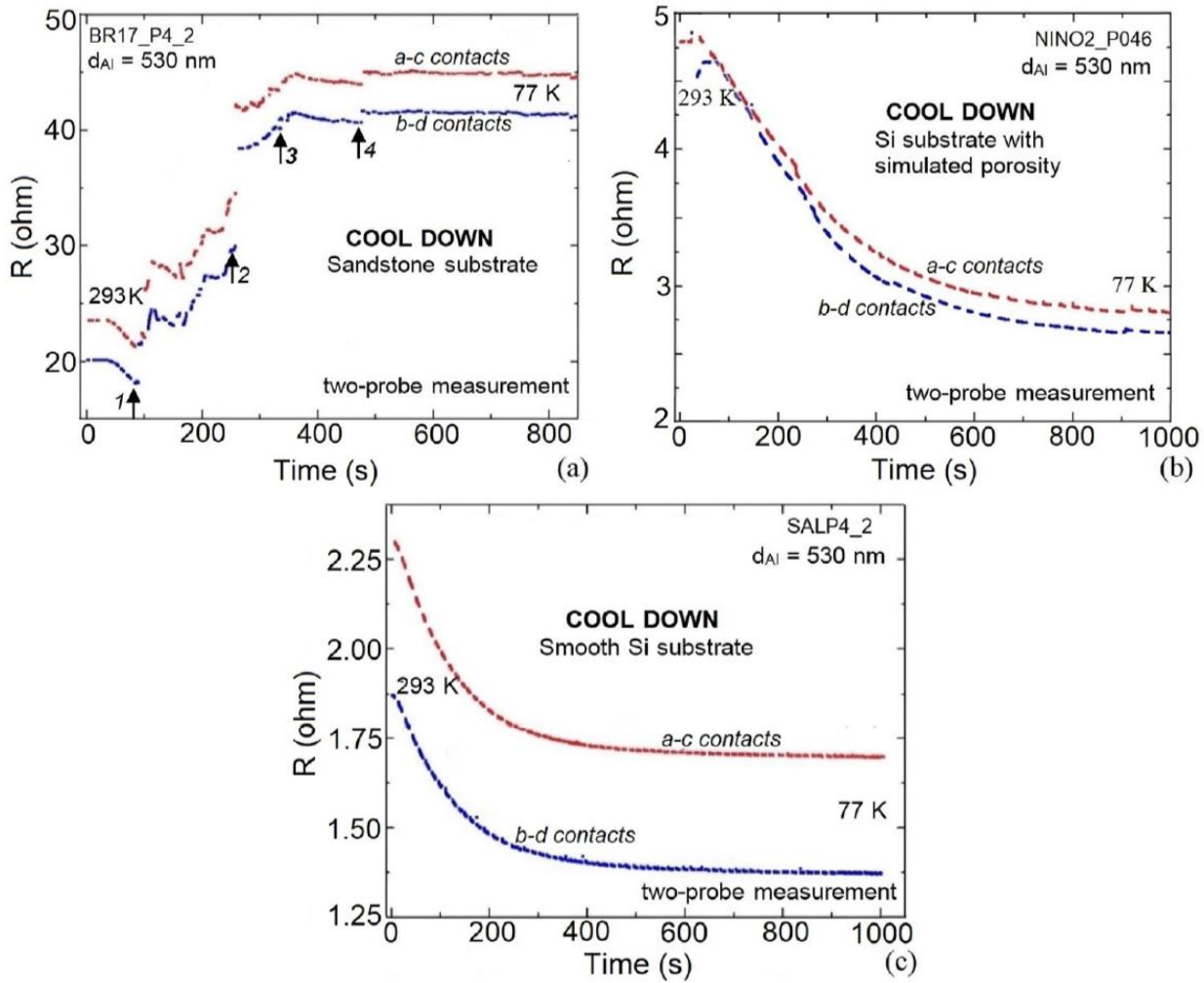

*Figure 4. Time-dependence of the resistance during a standard cool-down for samples with different substrates: (a) sandstone; (b) Si with simulated pores and surface roughness, and (c) smooth, flat Si. In the cool-down, the temperature decreased monotonically from 293 K (room temperature, which corresponds to t = 0) to near 77 K. At longer times, ~ 900 s, the temperature approached the boiling point of liquid nitrogen very closely but slowly, because during the cool-down process temperature difference decays with time exponentially. Thus, the sample's temperature never reached the precise temperature of liquid nitrogen and, thus, some insignificant changes in the resistance occurred at even longer times, say t ~ 900 s. Because the cool-down process is monotonic, the lowest temperature was achieved at the maximum time measured.*

After a cool-down experiment (~1000 s), a warming up process accompanied by a continuous resistance measurement was carried out, as shown in Figure 5. For the sandstone, the resistance initially increased by about 20% without any jumps, up to 175 s. This initial stage corresponds to low temperatures at which the coefficient of expansion of the sandstone is low and, thus, no significant displacement of sand grains occurred. The first jump was observed on the sandstone sample at $t \sim 200$ s; see Figure 5(a). The resistance exhibited a sharp jump downward (the first vertical arrow in the figure). The downward jump was unexpected because it represented a healing event, in which some weak links that were broken in the cool-down process were reestablished or reconnected. This jump might indicate that the shifts of the sand grains can be reversible. In the Si substrate with simulated roughness, Figure 5(b), and the flat, polished Si substrate in Figure 5(c), the resistance increased monotonically with increasing temperature. Such an increase is natural for a metallic film because the probability of electron–phonon-scattering is enhanced at higher temperatures.

Note that although the Si substrate with simulated roughness (blue curve in Figure 5(b)), exhibited some fluctuations in the resistance, the timing did not coincide with that of the other pairs of contacts. Thus, such fluctuations represent changes in the electrical contacts between the Al film and the signal wires (gold wires) attached to the film by In dots to measure its resistance. In fact, the four-probe measurements (see below) did not indicate such jumps because 4-probe measurements are not sensitive to variations in the contact resistance.

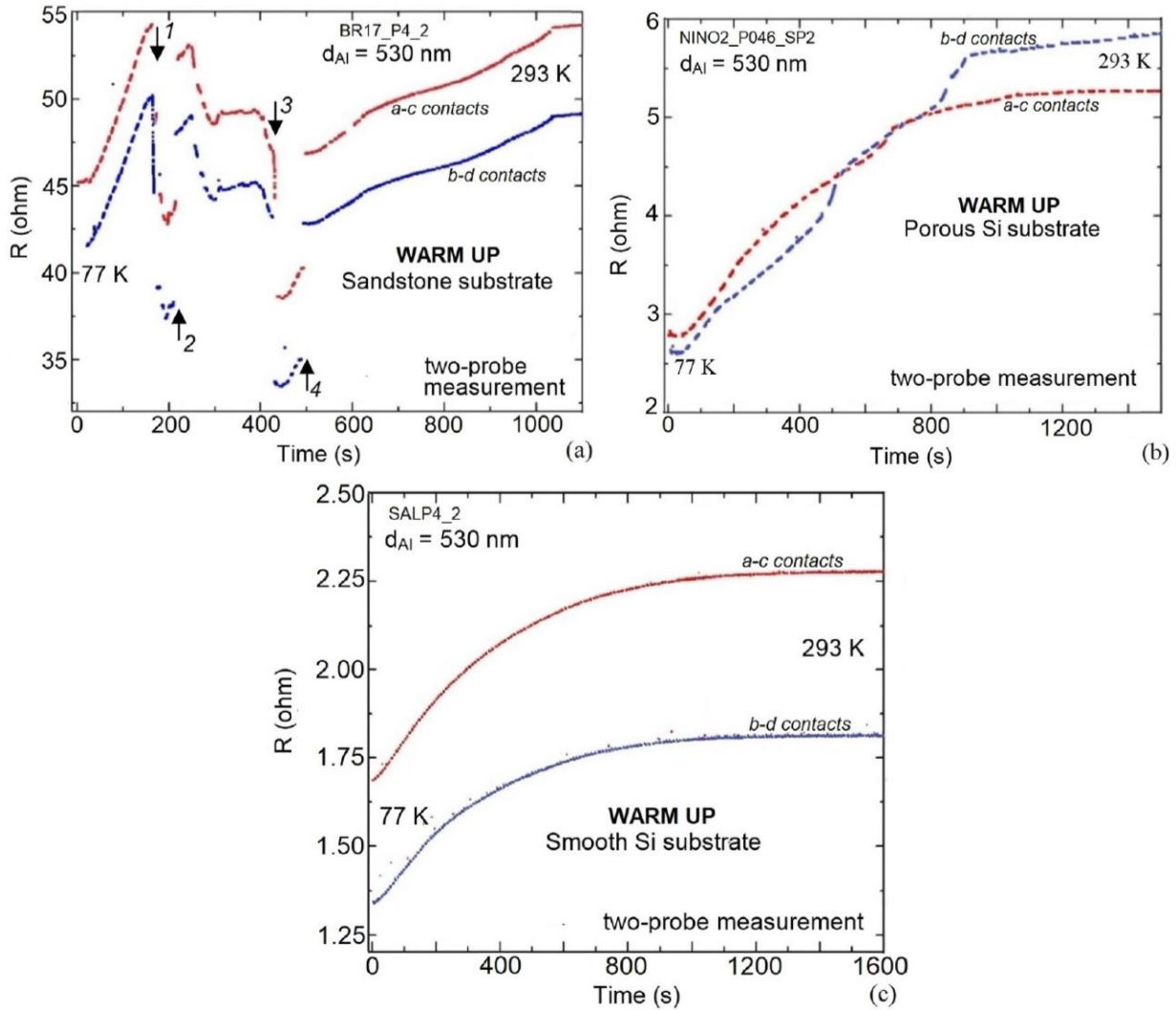

*Figure 5. Time-dependence of the resistance in the warm-up process for samples with various substrates: (a) sandstone; (b) Si with simulated surface roughness, and (c) smooth, flat Si.*

Note that Fig.4a exhibits a continuous increase of the resistance with cooling or, equivalently. While we do not have a firmly established explanation for this, we speculate that there may be microscopic weak links in the Al film that can be distorted due to small shifts of the grains, but the shifts are not to the extent that they can cause jumps. Note that if a small bridge's conductance is lower than the noise level of the setup, then, even if such a bridge is broken, it would not produce a visible jump in the conductance or resistance.

Figure 6 presents the true four-probe measured resistance during cooling and warming for the three types of substrates. The resistance is plotted versus time, assuming the temperature was changing monotonically with time. The data exhibit the same type of dependence on cooling or warming as did the two-probe measurements. The resistance of the sample with sandstone substrate changed by jumps during both cool-down and warm-up. In contrast, the resistance of the Al films on both the rough and smooth single-crystal Si substrates changed monotonically in both processes. Namely, the resistance decreased when the temperature was reduced and increased when the temperature increased.

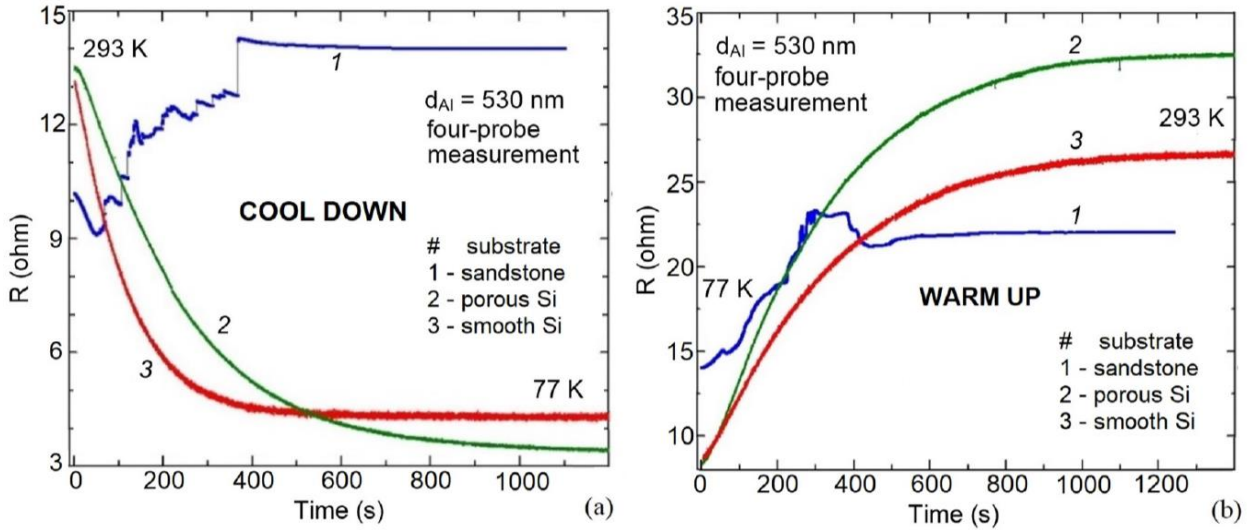

*Figure 6. Time-dependence of the resistance in the (a) cool-down, and (b) warm-up processes for samples with various substrates: 1: sandstone (BR18_P4); 2: Si with simulated surface roughness (NINO2_P046), and 3: smooth, flat Si (SALP4_2). (a) For comparison and clarity, the resistances of samples 2 and 3 were multiplied by 20 and 50 correspondingly. (b) A warming-up process. The Al film resistance is plotted versus time. The resistance of samples 2 was multiplied by a factor of 50 and the resistance of the sample 3 was multiplied by 100 in this plot.*

The sensing element in these experiments is the Al film, which is deposited by thermal evaporation on the sandstone surface. Our assumption is that the Al film consists of two qualitatively different regions. First, the film's islands that are deposited on individual sand grains and are well connected to the substrate, namely, the surface of the sand grains, and are stable as the temperature changes. This is confirmed by our test measurements of the Al film deposited on the Si substrates, which has a similar surface. The second type of region consists of the microscopic Al bridges which connect the Al film's islands. The bridges were confirmed to be narrow based on our observation of the microwave-induced quantum steps in the superconducting regime. Each bridge is smoothly connecting two neighboring sand grains. If the grains do not shift significantly, the bridge is stable. This was confirmed by our previous extensive measurements on such and similar metallic bridges [29]. Moreover, we report here a test measurement on a substrate with significant surface roughness, which is not composed of sand grains but, rather, of a single crystal of Si. It undergoes smooth deformations over cooling but does not generate jump-like local deformations. Correspondingly, such samples do not exhibit any jumps. Thus, we conclude that, overall, smooth changes of the sample do not contribute to the observed resistance jumps, which can occur only if two sand grains shift with respect to each other over a distance comparable to the film thickness, which is a fraction of a micrometer. Only such shifts can break an Al bridge that connects two Al islands resting on two neighboring sand grains (see Fig. 1).

The weak links act as sensors that can detect minute nanometer-scale displacements of the sand grains. Such displacements happen as the temperature of the sample is varied and the internal distribution of strain is changed. In the presence of metastable states such perturbations lead to sudden restructuring events, which shift the grain and break the local "sensors," i.e., the Al nanoscale weak links or nanobridges (See figure 1). In the initial cool-down experiment the number of events per unit surface per one cooling process was about 5 mm$^{-2}$. The number of events per unit surface in the subsequent warming-up process was about 3 mm$^{-2}$. Typical rate of the observed events was around 0.125 s$^{-1}$.

Thus, according to our picture, as the sandstone grains are displaced with the variations in the temperature and the resulting induced strain, some of the metastable states are relaxed, the nanobridges are broken, reducing the connectivity of the islands. It is known from percolation theory that as the local connectivity of conducting particles is lost, the resistance of the sample increases. When there is no longer a sample-spanning connected path of the particles – the islands in the present study – across the sample, the conductivity vanishes in an infinitely-large sample – the resistance diverges – or attain a small constant value beyond which it does not change anymore, implying that the resistance has reached its plateau. The jumps are caused by two factors. One is the finite size of the sample. Each time a nanobridge is broken, it has a finite effect on the conductivity and, hence, the resistivity. The jumps should reduce in size if the size of the sample increases significantly. The second factor is the nature of the rupture of the nanobridges, which is like micro- or nanofracturing process, and it is known that rupture is similar to a first-order transition that causes finite jumps in the conductivity or resistance, rather than a continuous increase or decrease in the conductivity or resistivity.

We note that our method is sensitive only to the rearrangements of the grains on the surface. The bulk rearrangements would not be detected, unless they cause a significant surface disturbance. Such rearrangements might release enough energy to initiate avalanche-type global rearrangements if critical states are present. We speculate that (1) each micro-seismic event begins from a single grain, and (2) that at least some abrupt shifts of sand grains should cause additional shifts of their neighbors and, thus, produce global rearrangement fracturing events. The second assumption is linked to the fact that critical states are present at least in some sandstones [30,31].

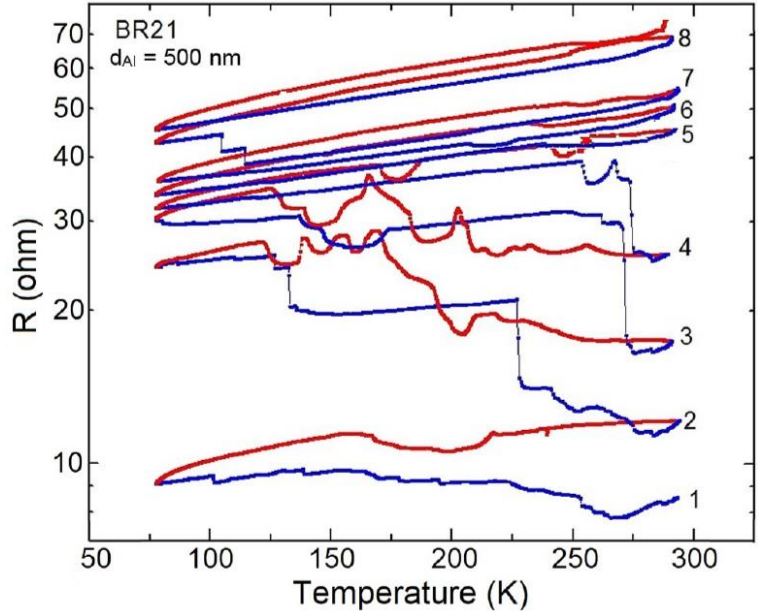

*Figure 7. Temperature-dependence of the resistance of Al film (530 nm). The measurement began at point 1. The resistance decreases from room temperature to the temperature of liquid nitrogen (blue curve). Note that the blue and red curves represent, respectively, the cooling and the warming stages. Jumps are present, and the resistance tends to increase (on average) each time the temperature was reduced in the three cycles. As the thermocycling of the sample continued, the jumps disappeared almost completely after cycle 5 and beyond, indicating that metastable regions on the sample relax, and no new metastable states occur.*

Figure 7 presents temperature-dependence of the resistance of the Al film deposited on sandstone. The cooling processes are shown as blue curves, and the heating processes are shown as red curves. The upward jumps in the resistance have developed, as discussed above, as the temperature was reduced from approximately 300 K to 77 K. There are five clear jumps in the first cool down process. The second cool down generates at least a couple of large jumps. We explain this by observing that the sandstone becomes weaker with thermal cycling, and some deep lying metastable state can relax, causing the large jumps. The process we report depends on the number of the cooling-warming cycles. The number of events of the cool-down process in the first 8 cycles was 10, 9, 5, 4, 0, 0, 2, 0. Roughly speaking, in the first three cycles, the resistance tended to increase with cooling and to decrease somewhat with warming. Our interpretation is that such trends represent some microscopic shifts of the sand grains from their initial positions that are metastable, i.e., not completely stable because of "frozen" internal stress, which is released when the sandstone contracts as it cools to cryogenic temperatures. Similar results were reported by Cai et al. [32]. As the surrounding temperature decreases, the sand grains shrink and deform, resulting in a reduction in the number and volume of the pores. In the warm-up cycle, on the other hand, some sand particles shift back to their original positions, and the resistance may decrease again, reveling some reversibility in the system. Yet, any additional shifts of the sand grains cause additional breaks of the weak links that provide continuity of the thin Al film, so the resistance can also increase in certain situations even in the warm-up process.

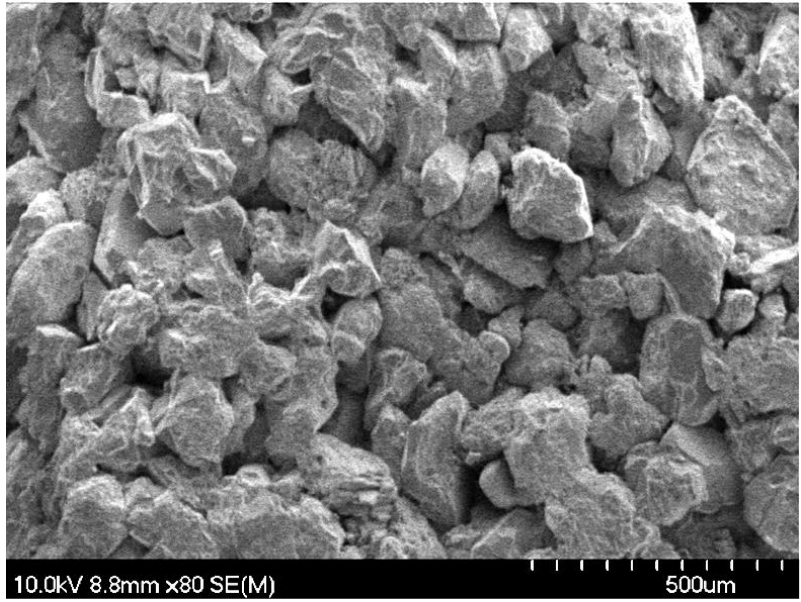

*Figure 8. A micrograph of the sandstone surface obtained by a scanning electron microscope. The surface was coated with a 20 nm thick film of AuPd.*

A scanning electron microscope (SEM) imaging of the sample's surface (Fig. 8) leads to an estimate of ~80 sand grains on the surface. As the surface is coated with an Al film, the grains are expected to be well covered, but the gaps between the grains might be connected by weak links, i.e., the small Al bridges that can be easily broken if a sand grain shifts from its original position. The number of such Al nanobridges is also estimated to be about 80 mm$^{-2}$, i.e., the same as the number of grains. In our model we estimate the probability of the sand grains shift as the number of the resistance jumps divided by the number of the grains, namely, $\rho = N_e/N_g$. Experimental results show that this probability is about 6% and 3% for the first cool down and warm up processes, respectively. The thermal cycling causes the events of grains rearrangements to becomes less probable, which means that some sort of "training" of the sample takes place. The sample is never disintegrated during the thermal cycling experiments, indicating that it does not produce macroscopic cracks of the size of the sample, i.e., the state of the sandstone is not critical.

After many cycles, all metastable configurations relax, and the sample enters a stable regime. In this case, the temperature variations of the film's resistance are only those natural for stable metallic samples in which the resistance drops with cooling and increases with heating in a monotonic fashion, i.e., without jumps; see curves 8 in Figure 7.

To confirm the presence of the nanometer-scale weak links in the Al film, we measured V-I characteristics of a sample at a very low temperature, $T \sim 300$ mK, at which Al is fully superconducting. The basic principles of such measurements are summarized in Ref. [27]. The resulting V-I curve is shown in Figure 9(a). The critical current, i.e., the current at which superconductivity is destroyed, was in the range 10 - 100 micro-amps, depending on the strength of the applied magnetic field. Previously, Bae et al. [29] measured the critical current as being approximately 100 micro-amps for a nanowire with a cross section of 1 μm$^2$, consistent with the critical current of our Al films deposited on sandstones. Thus, the total cross section of the weak links was approximately 1 μm. The number of weak links is estimated to be approximately 20, based on Figure 2(a), for example, as we approximate the number weak links by the number of sand grains. Thus, the cross section of each weak link was approximately ~0.05 μm$^2$. Therefore, the diameter of the weak links was roughly 250 nm, which is comparable to the coherence length of Al thin films [29]. Now, if the dimensions of weak links were comparable to the coherence length, each of them would act as a weak link, also called Dayem bridge, which would be expected to exhibit quantized voltage plateaus, if it is subjected to microwave radiation [29]. Such quantized plateaus were indeed observed in our samples (Fig. 9). To obtain the plot in Figure 9b, we measured many V-I diagrams with various frequencies of radiation. Then, we analyzed each diagram separately to determine the average height of the voltage steps. Then, all the results were collected in a single plot shown in Figure 9b. The steps follow the quantum relationship, $2e\Delta U = nhf$, where $n$ is an integer, $h$ is Planck's constant, and $f$ is the frequency of the applied electromagnetic radiation, e is the electronic charge on ΔU is the magnitude of the voltage steps. This equation represents a lock-in effect between the applied oscillating electromagnetic field and the phase rotation of the superconducting condensate wave function on the weak link. Such remarkable quantization of the voltage plateaus indicates that the size of the weak links is comparable to the coherence length, hence confirming that the typical dimensions of the weak links were at the nanometer-scale, and that they could be broken by small shifts of the sand grains on the nanometer length scale, as conjectured earlier.

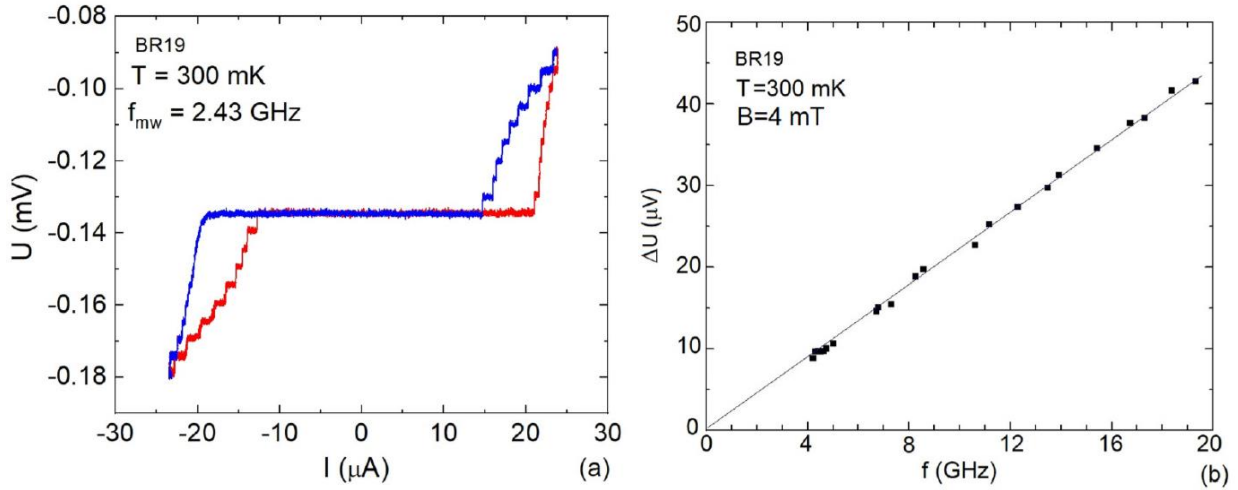

*Figure 9. (a) Voltage-current diagram of the Al film deposited on sandstone in the presence of microwave radiation. The microwave-induced steps (the analogue of the Shapiro steps in superconductivity) are related to a lock-in process between the applied electromagnetic field and the phase of the superconducting condensate wave function on the weak links. Such steps were not visible without microwaves (not shown). Red curve represents increasing and blue curve represents decreasing of the bias current. (b) Test of the quantization (Planck) formula E=2eΔU=hf, where E is the energy of superconducting electronic pairs, e is the electronic charge, ΔU – voltage drop on a weak link, h – Planck's formula, f – microwave frequency. The data are shown by the dots. The straight line represents the quantization formula (see the text). The fact that the height of the steps scaled linearly with frequency provides strong evidence that the electronic transport in the Al film was dominated by nanoscale weak links, i.e., the nanobridges.*

To establish whether evaporation of the Al film or the Al material itself was responsible for the observed anomalous behavior in the films deposited on sandstone, we also studied qualitatively different films. Specifically, a NbTi metallic film was deposited by magnetron sputtering on Berea sandstone. Note that thermal evaporation, used to deposit our Al films, is highly directional and produces shadow regions not covered by metal. Unlike thermal evaporation, magnetron sputtering is an isotropic deposition method, which covers all features of the sample surface rather homogeneously. Figure 10 presents the result of a four-probe measurement of electrical resistance of the NbTi film, deposited on Berea sandstone. This resistance versus time curve manifests the same trend as the thermally evaporated Al films discussed above. Namely, we observe abrupt jumps in the resistance of the NbTi film. As with Al films, we interpret the jumps as abrupt and pronounced shifts of some sand particles from their metastable quasi-equilibrium positions. Such shifts destroy local nanobridges in the metallic film - NbTi in this case - and cause the resistance to increase. We also observe a tendency of the resistance to increase when the temperature is reduced, which is unusual for metals. This effect is probably due to small micron scale shift of the sand grains, which produce a large number of very small microcracks in the NbTi film.

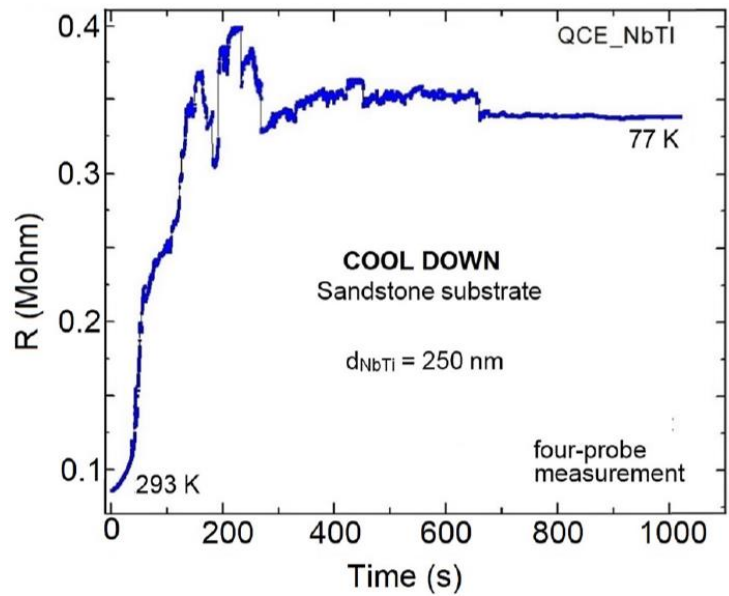

*Figure 10. Dynamic evolution of the resistance during cooling down of a NbTi film deposited on Berea sandstone. The film was deposited by magnetron sputtering technique, with the vacuum being $4 \times 10^{-10}$ Torr, while the rate of deposition was 0.7 Å/s. The thickness of the film was 250 nm.*

## V. ACOUSTIC MEASUREMENTS

In order to independently confirm that the observed resistance jumps are indeed caused by the sandstone substrate and not by the Al film itself, we carried out independent experiments using acoustic emission measurements. The sample is shown in Figure 11, with a microphone attached to the right face of the sample. The other side of the sample, having no microphone, was immersed in liquid nitrogen for the cool-down part of the cycle. Two cooldown-warmup cycles were carried out, each lasting about 160 s. Acoustic signals were converted to electrical signals by the piezo-microphone and were streamed continuously to our USB scope at 1 MHz. The piezo-transducer (the microphone) was mounted at one end of the sandstone brick that always remained at room temperature, while the opposite end of the sample was immersed in liquid nitrogen down to half sample length for about 80 s, and then removed and warmed up in air for about the same amount of time.

The results are presented in Figure 12. One can see very sharp spikes in the signals, which are not representative of the continuous nitrogen-boiling weak noise, but are probably due to discrete fracturing events of the sandstone itself. Moreover, the sharp spikes in the signals not due to the Al film breaking, since no Al (or any other) film was used in these experiments. In general, our expectation is that the acoustic events originate from the entire volume of the sample, whereas electrical events originate from the sample surface. Thus, a one-to-one correspondence is not expected. These observations confirm again that metastable microstates are present in sandstone and they relax as thermo-cycling is performed.

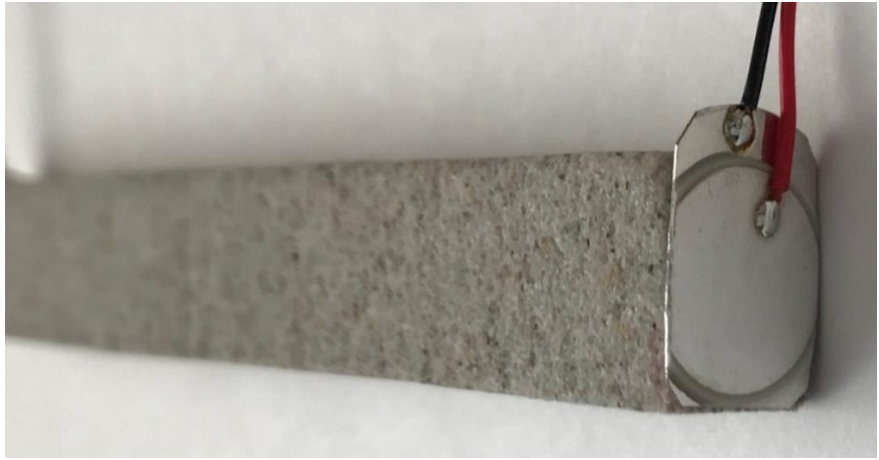

*Figure 11. Sandstone sample with a piezoelectric microphone attached to one end. The microphone is used to detect the emission of sound produced by the abrupt displacements of the sand grains, such as those we assign to the observed jumps. The sample contains no Al film or any metallic film coating.*

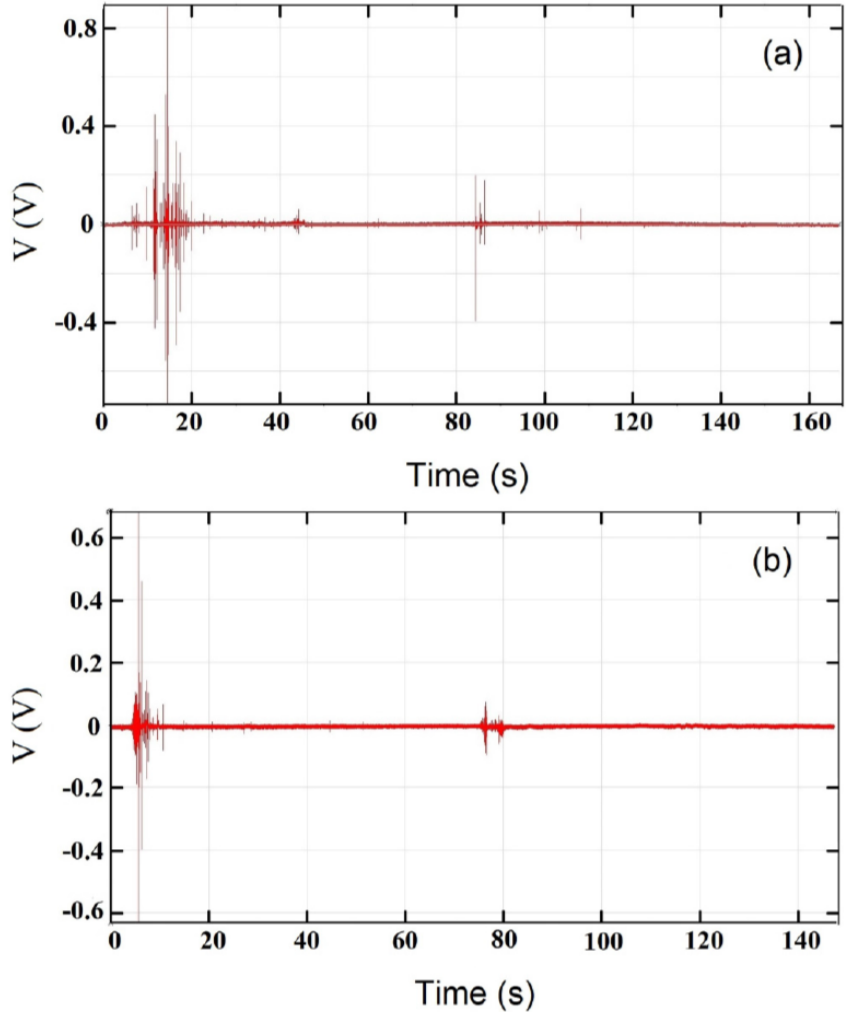

*Figure 12. Sound emission during a cool down (a) and the subsequent warming up (b) processes.*

Interestingly, we also observed sound burst events on the warming up part of the cycle and, notably more of them in the second thermal cycle, suggesting some slip-stick motion may be happening along the cracks formed during the thermal cycle.

## VI. SUMMARY

We observe metastable states in Berea sandstone samples. It is found that these states can relax upon cooling samples to cryogenic temperatures. The technique employed is based on metal (Al) deposition of thin films on the sample surface. Through superconductivity measurements, the resulting films have been established to be highly granular, with the grains connected by weak links or nanobridges. The weak links behave as Al nanowires, and act as "sensors" to small displacements of the surface sand grains. The film resistance exhibits unusual sharp jumps as the temperature is reduced. Moreover, the weak links can be reestablished when the temperature is increased. Such behavior suggests that sand particles forming the sandstone undergo abrupt shifts as the overall dimensions of the porous medium change with temperature. Such shifts in the position of the sand particles are unpredictable and reflect the presence of metastable states with concentrated mechanical energy, that are related to a particular realization of the mechanical structure of the sandstone. A control Si substrate with a rough, etched surface does not manifest such jumps, suggesting that the instability is essentially related to the granularity, and not the surface roughness of the sandstone. Another control experiments, carried out with a sensitive microphone attached to a sandstone sample (not coated with Al or anything else), indicate bursts of sound vibrations during a cool-down and a warm-up, indicating again that metastable states are present.

## ACKNOWLEDGMENTS

This work was supported as part of the Center for Geologic Storage of $CO_2$, an Energy Frontier Research Center funded by the U.S. Department of Energy, Office of Science, Basic Energy Sciences under award number DE-SC0C12504.